\documentclass[twocolumn,showpacs,prl,10pt]{revtex4}
\usepackage[english]{babel}
\usepackage{amsmath,amssymb}
\usepackage{times}
\usepackage{epsfig}

\begin{document}

\title {Scaling of the magnetic response in doped antiferromagnets}
\author{P. Prelov\v sek$^{1,2}$, I. Sega$^1$, and J. Bon\v ca$^{1,2}$}
\affiliation{$^1$J.\ Stefan Institute, SI-1000 Ljubljana,
Slovenia}
\affiliation{$^2$Faculty of Mathematics and Physics, University
of Ljubljana, SI-1000 Ljubljana, Slovenia} 
\date{\today}

\begin{abstract}
A theory of the anomalous $\omega/T$ scaling of the dynamic magnetic
response in cuprates at low doping is presented. It is based on the
memory function representation of the dynamical spin suceptibility in
a doped antiferromagnet where the damping of the collective mode is
constant and large, whereas the equal-time spin correlations saturate
at low $T$. Exact diagonalization results within the $t$-$J$ model are
shown to support assumptions. Consequences, both for the scaling
function and the normalization amplitude, are well in agreement with
neutron scattering results.

\end{abstract}

\maketitle

Magnetic properties of cuprates, as they evolve by doping the
reference antiferromagnetic (AFM) insulator and lead to a high-$T_c$
superconductor, have been so far a subject of intensive experimental
and theoretical investigations.  One of the puzzles
awaiting  proper theroretical explanation is the scaling behavior
of the magnetic response observed in cuprates, mostly in the regime of
low doping \cite{kast}. It has been first found by the inelastic
neutron scattering experiments in La$_{2-x}$Sr$_x$CuO$_4$ (LSCO) at
low doping $x=0.04$ \cite{keim} that ${\bf q}$-integrated  spin
susceptibility $\chi_L''(\omega)$ follows a simple universal behavior
in terms of the scaling variable $\omega/T$. Similar scaling has been
observed also in La$_{2-x}$Ba$_x$CuO$_4$ \cite{hayd}, in 
 YBaCu$_3$O$_{6+x}$ (YBCO) with $x=0.5, 0.6$
\cite{birg}, and even more pronounced in Zn-substituted YBCO \cite{kaku}.

In particular, experiments on cuprates at low doping indicate that one
can represent results for the local susceptibility in a broad range of
$\omega$ and $T$ as $\chi_L''(\omega,T)= I(\omega) f(\omega/T)$ where
$I(\omega)=\chi_L''(\omega,T=0)$. The scaling function should approach
$f(x \to \infty)=1$ and the simplest form invoked in the analysis is
$f(x) \sim (2/\pi){\rm tan}^{-1}(Ax)$.  It seems, however, that the
function is not universal for all cases, i.e., $A \sim 1.0 - 2.2$
varies between YBCO results \cite{birg,kaku} and LSCO, whereby in the
latter case corrections to the simplest form reveal an even better
agreement \cite{keim}. At the same time it is found that the inverse
AFM correlation length $\kappa=1/\xi$, as extracted from the ${\bf
q}$-dependent $\chi''({\bf q},\omega)$, saturates at low $\omega$ and
$T$.  The largest response is at the AFM wavevector ${\bf
Q}=(\pi,\pi)$ and as a consequence of the scaling  the peak in
$\chi''({\bf Q},\omega)$ should move downward with decreasing $T$,
this being in fact established in YBCO \cite{regn}.  It should be
noted that also NMR relaxation experiments test and confirm the
$\omega/T$ scaling of $\chi_L''(\omega)$ at $\omega \to 0$
\cite{imai}.

Such a $\omega/T$ scaling is inconsistent
with the concept of usual Fermi liquid. This has been recognized quite
early and the concept of the 'marginal' Fermi liquid has been
introduced \cite{varm} to explain scaling of the magnetic response as
well as of other anomalous electronic properties. One appealing
explanation still considered is the vicinity of the quantum critical
point \cite{chak}. However, the latter should in general require also
a critical variation of $\kappa(\omega,T)$, as indeed observed in LSCO
near the optimum doping \cite{aepp}. A random-phase-approximation
treatment of $\chi({\bf q},\omega)$ \cite{moriy} yields a relaxation
rate $\Gamma \propto 1/\xi^2$, and the scaling form could be reproduced
provided  that the correlation length is critical, i.e., $\xi
\propto T^{-1/2}$, which is not consistent with experiments
\cite{keim,kast}.  On the other hand, numerical investigations of the
twodimensional $t$-$J$ model confirm the scaling of
$\chi''_L(\omega)$ \cite{jakl}, although results are restricted to
rather high $T$ as compared to experiments.

In the following we will argue that the anomalous $\omega/T$ scaling
of the magnetic response can be understood as a consequence of few
simple ingredients which appear to be valid for doped AFM in the
normal state: a) the collective mode is strongly overdamped, whereby
the damping is nearly $\omega$- and $T$- independent at low $\omega$,
and b) there is no long-range spin order at low $T$, so that static
spin correlations saturate with a finite $\xi$. It will be shown that
these prerequisites are sufficient to reproduce several experimental
findings for $\chi_{L}(\omega)$.

Within the memory function approach \cite{mori} the dynamical spin
susceptibility $\chi_{\bf q}(\omega)=-\langle \!\langle S^z_{\bf
q};S^z_{\bf q} \rangle \!\rangle_{\omega}$ can be expressed in the
form
\begin{equation}
\chi_{\bf q}(\omega)=\frac{-\eta_{\bf q}}{\omega^2+\omega 
M_{\bf q}(\omega) - \omega^2_{\bf q}}\,, \label{chiq}
\end{equation}
suitable for the analysis of the magnetic response, as manifest in
underdoped AFM \cite{sega}. $\omega_{\bf q}$ represents
the frequency of a collective mode provided that the mode damping is
small, i.e., $\gamma_{\bf q}\sim M^{\prime\prime}_{\bf q} (\omega_{\bf
q}) <\omega_{\bf q}$. For $\gamma_{\bf
q}>\omega_{\bf q}$ the mode is overdamped. The advantage of the
form (\ref{chiq}) is that it can fullfil basic sum rules even for an
approximate $M^{\prime\prime}_{\bf q}$. Thermodynamic quantitites
entering Eq.~(\ref{chiq}) can be expressed as
\begin{equation}
\eta_{\bf q}=-{\dot\iota}\langle [S^z_{-\bf q}\, ,\dot{S}^z_{\bf q})]\rangle\,,
\quad \omega^2_{\bf q}=\eta_{\bf q}/\chi^0_{\bf q} \,,
\end{equation}
where $\chi^0_{\bf q}=\chi_{\bf q}(\omega=0)$ is the static susceptibility.

$\eta_{\bf q}$ is closely related to the spin stiffness and can be
expressed in terms of the static correlation functions, and is expected
to  be weakly ${\bf q}$-dependant for ${\bf q} \sim {\bf Q}$.  Static
$\chi^0_{\bf q}$ (or $\omega_{\bf q}$) remains to be determined, even
for known $M_{\bf q}(\omega)$.  Instead of directly evaluating
$\chi^0_{\bf q}$, being quite a sensitive quantity, we rather fix it
by the sum rule
\begin{equation}
\frac{1}{\pi}\int_0^\infty d\omega ~{\rm cth}\frac{\omega}{2T}
\chi^{\prime\prime}_{\bf
q}(\omega)= \langle S^z_{-{\bf q}} S^z_{\bf q}\rangle = C_{\bf q}\, ,
\label{eqsum}
\end{equation}
given in terms of equal time  correlations, which are expected
to be less $T$-dependent. $C_{\bf q}$
are bound by a local constraint $(1/N)\sum_{\bf q} C_{\bf q} =
(1-c_h)/4$, where $c_h$ is an effective hole doping.

Let us define now our central assumptions. We first take that static
correlations follow the standard Lorentzian form, i.e., $C_{\bf
q}=C/(\kappa^2+\tilde q^2)$ \cite{keim}, where $\tilde {\bf q}={\bf q}-{\bf
Q}$, although our results are not very sensitive to the explicit form of
$C_{\bf q}$ (at fixed $\kappa$). $\kappa$ is taken as a $T$-independent
constant,  at least on approaching  low $T$. Such an assumption for $\kappa$
is  consistent with the neutron scattering data for weakly doped LSCO
\cite{keim}  and YBCO \cite{birg,kaku},  as well as with results for the
$t$-$J$  model at finite doping  \cite{sing}. It furthermore
indicates  that the system  remains paramagnetic with  finite AFM $\xi$ down
to  the lowest $T$, as  well as the absence of any ordered ground state.

Less plausible is the second assumption that the damping is also
constant, $M^{\prime\prime}_{\bf q}(\omega) \sim \gamma$, i.e.,
(roughly) independent of $\omega$, $\tilde {\bf q}$ and $T$, or at
least not critically dependent on these variables. We can give several
arguments in favor of this simple choice. Recently the present authors
\cite{sega} studied the spin dynamics within the $t$-$J$ model,
\begin{equation}
H=-\sum_{i,j,s}t_{ij} \tilde{c}^\dagger_{js}\tilde{c}_{is}
+J\sum_{\langle ij\rangle}({\bf S}_i\cdot {\bf S}_j-\frac{1}{4}
n_in_j) \, , \label{eq1}
\end{equation}
with the nearest neighbor $t_{ij}=t$ and next-nearest-neighbor hopping
$t_{ij}=t'$. It has been shown that the dominant contribution to the
damping $M^{\prime\prime}_{\bf q}(\omega)$ in a doped system comes
from the decay of spin fluctuations into fermionic electron-hole
excitations \cite{sega}. If the fermionic excitations in a doped
system behave as in a Fermi liquid, and the Fermi surface crosses the
AFM zone boundary, the damping in the normal state is essentially
constant at low $\omega$, and also weakly dependent on $T$ and ${\bf
q} \sim {\bf Q}$. This is clearly very different from an undoped AFM
where one expects vanishing $M^{\prime\prime}_{\bf q}(\omega_{\bf q})$
for ${\bf q} \to {\bf Q}$ and $T \to 0$ \cite{tyc}.

In order to support the simplification of constant $\gamma$, we
present in the following numerical results for the $t$-$J$ model,
obtained via the finite-$T$ Lanczos method (FTLM) \cite{jakl} for a
system of $N=20$ sites on a square lattice with periodic boundary
conditions. The model is analysed for the parameter $J/t=0.3$ as
appropriate for cuprates (note also the relevant value $t \sim
400~$meV), and within the regime of low hole doping, $c_h=N_h/N\leq
0.15$. Note that results within the FTLM have
macroscopic relevance for high enough $T$, while at $T<T_{fs}$ they
become influenced by finite-size effects.  As a criterion
for $T_{fs}$ we use the thermodynamic sum $\bar
Z(T)=\mathrm{Tr~exp}(-(H-E_0)/T)$ and the requirement $\bar Z(T_{fs})
= Z^*\gg 1$ \cite{jakl}.  In the cases discussed here $T_{fs}\sim
0.1~t$ at intermediate doping, i.e., $T_{fs} \sim 400~$K in terms of
cuprate parameters \cite{jakl}. Within the FTLM we calculate directly
$\chi^{\prime\prime}_{\bf q}(\omega)$. Since $\eta_{\bf q}$ and
$\omega_{\bf q}$ are given as frequency moments of
$\chi^{\prime\prime}_{\bf q}(\omega)$, it is then easy to extract also
the damping function $M_{\bf q}(\omega)$ via Eq.(\ref{chiq}).  

In Fig.~1 we present results both for $\chi^{\prime\prime}_{\bf
q}(\omega)$ and for $M^{\prime\prime}_{\bf q}(\omega)$, for fixed
doping $c_h=2/20$ and $T=0.15~t > T_{fs}$. In the
analysis a smoothing $\epsilon=0.07~t$ is used for convenience. One
can conclude that in the presented case we are clearly dealing with
overdamped spin dynamics for all presented ${\bf q}$. In spite of
widely different $\chi^{\prime\prime}_{\bf q}(\omega)$ the damping
function $M^{\prime\prime}_{\bf q}(\omega)$ is nearly constant in a
broad range of $\omega<t$ and almost independent of ${\bf q}$. 
For this particular $c_h$ we estimate $\kappa=0.7$ so that
the span of ${\bf q}$ goes beyond $\tilde q>\kappa$.

\begin{figure}[htb]
\centering
\epsfig{file=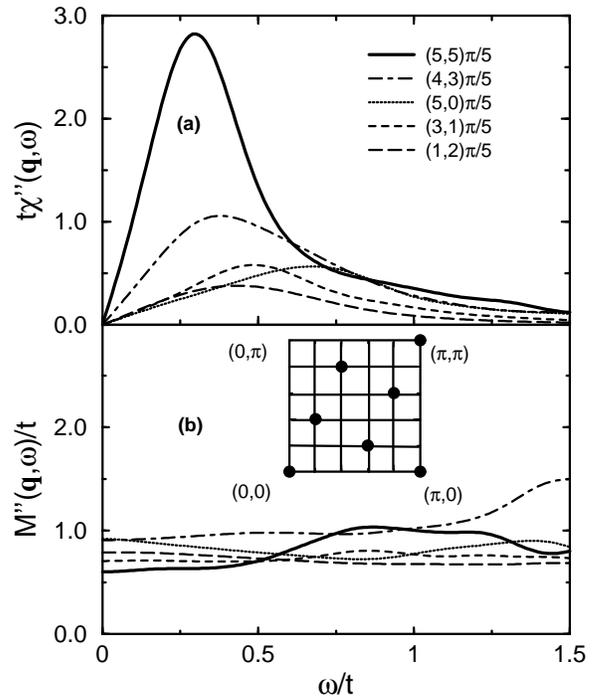,width=95mm,angle=-90}
\caption{(a) Spin susceptibility $\chi^{\prime\prime}_{\bf q}(\omega)$
within the $t$-$J$ model for doping $c_h=2/20$ and $T=0.15~t$, for
different ${\bf q}$, b) damping function $M^{\prime\prime}_{\bf
q}(\omega)$ for the same parameters. Inset shows nonequivalent ${\bf
q}$ for the lattice of $N=20$. }
\label{fig1}
\end{figure}

Fig.~1 confirms that it is meaningful to extract $\gamma_{\bf
q}=M^{\prime\prime}_{\bf q}(\omega=0)$, which we present in Fig.~2 for
${\bf q}={\bf Q}$ and various doping $c_h=N_h/N\leq 0.15$ as a
function of $T$. As expected, our results in an undoped system are
consistent with vanishing $\gamma_{\bf Q}(T \to 0)$ whereas for $c_h>0$ they
lead to a finite
extrapolated value $\gamma_{\bf Q}(T \to 0)$. Note that the slopes of
$d\gamma_{\bf Q}(T)/dT$  in Fig.~2 are quite similar for
all presented $c_h$. This can be interpreted as a signature that the
damping is a sum of the spin-exchange contribution and the fermionic
contribution \cite{sega}.  Only the spin-exchange term is active in an
undoped system and apparently it adds to the fermionic damping, the
latter dominating the $T\to 0$ behavior in a doped system. It should
be also noted that the characteristic (saturation) scale for the
dominant spin-exchange damping is $T\sim J$ which is far above the $T$
investigated in experiments. Hence, for the $T$ window of interest our
results confirm that in a doped system the simplification of constant
$\gamma$ is sensible. On the other hand, it is evident from Fig.~2
that $\gamma$ increases with doping, becoming very large $\gamma \sim
t$ on approaching the 'optimum' doping $c_h \sim
0.15$.
 
\begin{figure}[htb]
\centering
\epsfig{file=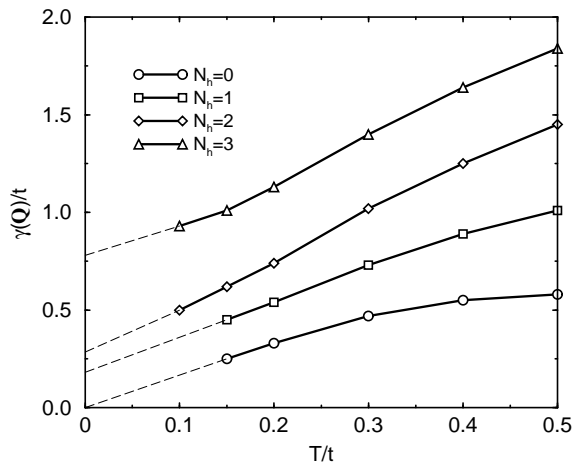,height=75mm,angle=-90}
\caption{Low-frequency damping $\gamma_{\bf Q}$ as a function of
$T$ for various doping $c_h=N_h/N$.}
\label{fig2}
\end{figure} 

Let us consider the consequences of proposed simplifications taking
both $\gamma$ and $\eta$ as constants. The dynamical susceptibility
now takes the resonance form
\begin{equation}
\chi''_{\bf q}(\omega)=\frac{\eta \gamma \omega}{(\omega^2 -
\omega^2_{\bf q})^2 + \gamma^2 \omega^2}\,, \label{chiim}
\end{equation}
which we have to investigate together with the sum rule,
Eq.(\ref{eqsum}). At given ${\bf q}$ there are several regimes with
respect to the values of $T$, $\gamma$ and $\omega_{\bf q}$.  Most
evident is the situation for $\gamma, T\ll \omega_{\bf q}$ with an
underdamped mode with a frequency $\omega_{\bf q} \sim \eta/(2 C_{\bf
q})=\alpha(\tilde q^2+ \kappa^2)$ where $\alpha=\eta/2C$.  In such a
case both $\chi''_{\bf Q}(\omega)$ as well as the local susceptibility
$\chi''_L(\omega)=(1/N)\sum_{\bf q} \chi''_{\bf q}(\omega)$ show more
or less (depending on $\gamma$) pronounced gap for $\omega<\omega_{\bf
Q} \sim \alpha \kappa^2$.  This regime clearly does not exhibit the
desired $\omega/T$ scaling.  On the other hand, even in a weakly doped
AFM at low $T$ with $\kappa \ll 1$ 
one should enter such an underdamped regime for the collective mode
with $\tilde q \gg \kappa$. Still, in this case for $\omega_{\bf q} >
\gamma$ one would expect that the dispersion becomes that of AFM
paramagnons with $\omega_{\bf q} \propto \tilde q$. This indicates
that a Lorentzian form for $C_{\bf q}$ presumably is not appropriate
for such a regime and should be modified for $\tilde q \gg \kappa$.

Experiments on cuprates as well as numerical results for the
$t$-$J$ model (as apparent in Figs.1,2), however, show that in the
normal state the collective mode is {\it always overdamped} in the
vicinity of ${\bf q}={\bf Q}$, i.e., $\omega_{\bf Q} < \gamma$.  Now,
one gets a simple Lorentzian for low $\omega<\omega_{\bf q}$,
\begin{equation}
\chi''_{\bf q}(\omega) \sim \frac{\eta}{\gamma} \frac{\omega}
{(\omega^2 + \Gamma^2_{\bf q}) } , \qquad \Gamma_{\bf q}=
\frac{\omega^2_{\bf q}}{\gamma}, \label{chiim1}
\end{equation}
and $\Gamma_{\bf q}<\omega_{\bf q}$. An overdamped form as in
Eq.(\ref{chiim1}) has been frequently invoked in the analysis of the
magnetic response \cite{moriy,keim} in the normal state of cuprates.
However, without the knowledge of $\omega_{\bf q}$ Eq.(\ref{chiim1})
is not sufficient to analyse the relation with the sum rule,
Eq.(\ref{eqsum}).

Let us first discuss low $T \to 0$. In this case the l.h.s. of
Eq.(\ref{eqsum}) can be explicitly integrated, and for $\omega_{\bf q}
< \gamma$ we get $C_{\bf q} \sim (2\eta/\pi\gamma)
\ln(\gamma/\omega_{\bf q})$. The relevant quantity is the peak
frequency $\omega_p=\Gamma_{\bf Q}(T\to 0)$. We see that the crucial
parameter is
\begin{equation}
\zeta = C\pi\gamma/(2 \eta \kappa^2), \qquad \omega_p \sim \gamma 
{\rm e}^{-2\zeta}, \label{zeta}
\end{equation}
which exponentially renormalizes $\omega_p$. Since $C$ is fixed by the
total sum rule, i.e., $C \sim (1-c_h) \pi /(2 \ln (\pi/\kappa)) \sim
O(1)$ and $\eta \sim 0.6~t$ \cite{sega} at low doping, $\zeta$ is effectively
governed by the ratio $\gamma/\kappa^2$. Our results for the $t$-$J$
model, as presented above as well as the analysis of experiments on
cuprates, indicate that generally $\zeta \gg 1$.

A nontrivial quantity which is the consequence of the presented $T=0$
analysis is the local $\chi_L''(\omega,0)=I(\omega)$ directly related
to the measured 'normalization' function \cite{keim,kaku}. In order to
evaluate the latter we first find for each $\tilde q$ the appropriate
$\omega_{\bf q}$ satisfying the sum rule (\ref{eqsum}) and then
integrate over ${\bf q}$. Results for $I(\omega)$ at various $\zeta$
are presented in Fig.~3. For convenience we fix $\gamma =0.2~t$, which
appears to correspond (see Fig.~2) to low doping $c_h\sim 0.05$, close
to doping in cuprates with observed scaling behavior. We note that the
range $\zeta = 2 - 8$ presented in Fig.~3 corresponds to $\kappa= 0.35
- 0.19$. We see from Fig.~3 that the behavior for all $\zeta$ is
qualitatively similar at high $\omega$ while the difference is mainly
in the position of $\omega_p$ where $I(\omega)$ is maximum.

\begin{figure}[htb]
\centering
\epsfig{file=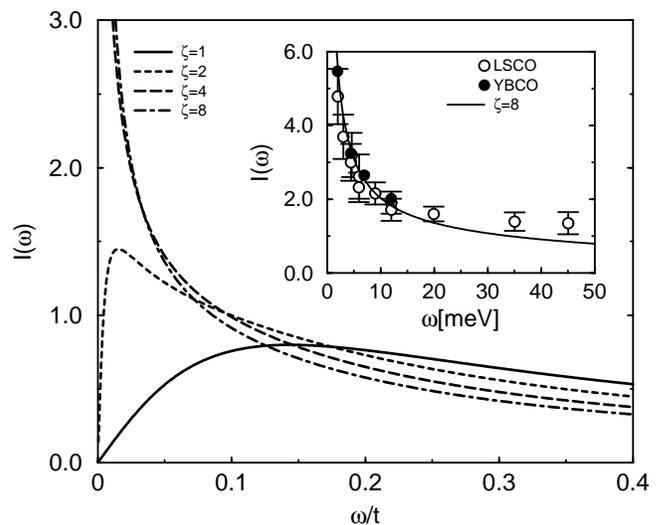,height=85mm,angle=-90}
\caption{ Local susceptibility $\chi''_L(\omega,T=0)$ for different
$\zeta$. The inset shows the comparison with (intensity scaled)
experimental data for 'normalization' function $I(\omega)$ in LSCO
\cite{keim} and Zn-doped YBCO \cite{kaku}. 
The vertical scales are adjusted since the experimental scales are not in
absolute units.} 
\label{fig3}
\end{figure}  

We can make a direct comparison with experiments on cuprates at low
doping which reveal a nontrivial $I(\omega)$, as presented in the
inset of Fig.~3. We first note that data for LSCO at $x=0.04$
\cite{keim} and Zn-substitued YBCO \cite{kaku} are quite similar. Both
indicate a steep increase of $I(\omega)$ below $\omega=10~$meV and no
sign of saturation even at $\omega=2~$meV. In terms of our analysis
this means that $\zeta \gg 1$.  The comparison of $I(\omega)$ with our
results at $\zeta=8$ (any $\zeta \gg 1$ would be in fact quite
satisfactory) reveals very good agreement . The difference seems to appear at larger $\omega>20~$meV which
could be again due to our Lorentzian form of $C_{\bf q}$. Namely,
taking for $\tilde q \gg \kappa$ $C_{\bf q} \propto 1/\tilde q$ would
lead to a flat $\chi_L''(\omega) \sim$~const., as in an ordered AFM
where only transverse fluctuations - magnons - contribute.

We next discuss the behavior at $T>0$. It is evident that for
$T>\omega_p$ the temperature dependence of $\Gamma_{\bf Q}(T)$ (or
$\omega_{\bf Q}(T)$) becomes crucial. In order to satisfy the sum rule
(\ref{eqsum}) it follows $\Gamma_{\bf Q}(T) \propto T$ which is the origin of the
$\omega/T$ scaling.  In Fig.~4 we present the 'scaling' function
$f(\omega/T)= \chi_L''(\omega,T)/\chi_L''(\omega,0)$ for various $T$
and chosen $\zeta=8$. Results confirm that indeed $f(\omega/T)$ is
nearly universal in a very broad range of $T$, i.e., between $T \sim
\omega_p$ and $T \sim \gamma$. We show in Fig.~4 for comparison also
experimental scaling function for Zn-substituted YBCO \cite{kaku}
which generally fits our results very well. It is also evident that at
least at lower $T<0.05~t$ our scaling function can be closely
represented by $f(x)=(2/\pi){\rm atan}(Ax)$ with $A \sim 1.2$.

There is still some dependence of $f(x)$ on parameter $\zeta$.  A
general tendency is that at larger $\zeta \gg 1$ we observe the
saturation at somewhat smaller $\omega/T \sim 1$, i.e., appropriate $A$
increases. On the other hand, for decreasing $\zeta \to 1$ we get
$f(x\to 0)>0$ and the saturation moves to somewhat higher $x$.

\begin{figure}[htb]
\centering
\epsfig{file=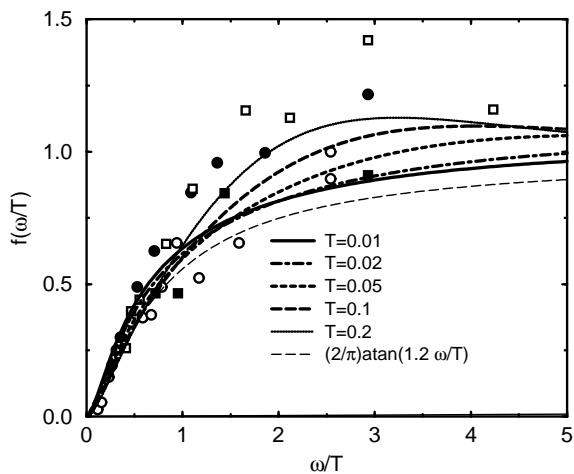,height=75mm,angle=-90}
\caption{ Scaling function $f(\omega/T)$ for $\zeta=8$ for different
$T$ (in units of $t$). For comparison data for Zn-substitued YBCO
are also plotted, as measured for different energies $\omega$, taken
from Ref.\cite{kaku}. }
\label{fig4}
\end{figure}  

In conclusion, we presented a theory giving an explanation for the
anomalous $\omega/T$ scaling behavior in magnetic response of doped
AFM. It is based on two key assumptions: a) the saturation of
static  spin correlations and of the correlation length $\xi$ at low $T$
and, b) on the constant damping $\gamma$ of the collective mode. While the
former is supported by experimental data \cite{keim,
birg, kaku}, the latter follows from numerical analysis on finite clusters
within  the $t$-$J$ model and deserves further theoretical
confirmation. However, both requirements are intimately related since they are
consistent with a paramagnetic liquid, with fermionic excitations
dominating low-$\omega$, low-$T$ behavior. This picture is 
picture is supported by ARPES experiments revealing   well pronounced
quasiparticle excitations even in a weakly doped LSCO \cite{yosh}.

In the presented picture the broad validity of the scaling is due to
large $\zeta \gg 1$, i.e., the AFM ${\bf q}={\bf Q}$ collective modes
are heavily overdamped even at low $T$. This is consistent both with
neutron scattering results in cuprates as well as with available
numerical results within the $t$-$J$ model. Note, however, that our
results should remain equally valid for the case of strong-coupling, i.e.,
$U\gg t$ effective single band Hubbard model, which in the low energy sector
reduces to the $t$-$J$ model with $J=4t^2/U$.
Our scenario for  the $\omega/T$ scaling differs from a quantum-critical one
\cite{chak}  in spite of a similar behavior of $\Gamma_{\bf Q} \propto T$,
since $\kappa$ does not scale in the same way. It should be pointed out,
however, that $\tilde\kappa$ as deduced, e.g., from $\chi^0_{\bf q}$ or from
$\chi_{\bf q}''(\omega)$ at fixed $\omega$ is significantly reduced, i.e.,
$\tilde\kappa<\kappa$ at low $T$. 

There are still some open questions. The theory predicts the existence
of the crossover temperature $T \sim \omega_p$ below which the scaling
would cease to exist, and the response would approach
$\chi_L''(\omega,T=0)$. Such a saturation has so far not been reported
for weakly doped LSCO and Zn-substitued YBCO, which do not exhibit
other phase instabilities at low $T$. One should, however, not forget
a possible influence of disorder, since the same region of the phase
diagram is often associated with the spin glass character
\cite{kast}. Since our assumptions appear to remain valid also at
higher (up to optimum) doping in the normal state, we can speculate on
a possible validity of the same scenario in this regime as well, provided that
other instabilities are absent (e.g., superconductivity, stripe
ordering).  The indication for the latter are the NMR $T_1$ relaxation
results \cite{imai} showing the same scaling in LSCO for $T>T_c$ up to
$x=0.15$.

Authors acknowledge the support of the Ministry of Education, Science
and Sport of Slovenia.


\begin{thebibliography}{99}
\bibitem{kast} for a review see M.\ A.\ Kastner, R.\ J.\ Birgeneau, 
G.\ Shirane, and Y.\ Endoh, Rev.\ Mod.\ Phys.\ \textbf{70}, 897 (1998).
\bibitem{keim} B.\ Keimer {\it et al.}, Phys.\ Rev.\ Lett. \textbf{67},
1930 (1991); Phys.\ Rev.\ B \textbf{46}, 14034 (1992).
\bibitem{hayd} S.\ M.\ Hayden {\it et al.}, Phys.\ Rev.\
Lett. \textbf{66}, 821 (1991).
\bibitem{birg} R.\ J.\ Birgeneau {\it et al.}, Z.\ Phys.\ B \textbf{87},
15 (1992); B.\ J.\ Sternlieb {\it et al.}, Phys.\ Rev.\ B \textbf{47},
5320 (1993).
\bibitem{kaku} K.\ Kakurai {\it et al.}, Phys.\ Rev.\ B \textbf{48},
3485 (1993).
\bibitem{regn} L.\ P.\ Regnault {\it et al.}, Physica B \textbf{213-4},
48 (1995).  
\bibitem{imai} T.\ Imai, C.\ P.\ Slichter, K.\ Yoshimura, and K.\ Kosuge,
Phys.\ Rev.\ Lett. \textbf{70}, 1002 (1993).
\bibitem{varm} C.\ M.\ Varma, P.\ B.\ Littlewood, S.\
Schmitt-Rink, E.\ Abrahams, and A.\ E.\ Ruckenstein, Phys.\ Rev.\ 
Lett. \textbf{67}, 1996 (1989).
\bibitem{chak} S.\ Chakravarty, B.\ I.\ Halperin, and D.\ R.\ Nelson,
Phys.\ Rev.\ Lett. \textbf{60}, 1057 (1988); Phys.\ Rev.\ B \textbf{39},
2344 (1989).  
\bibitem{aepp} G.\ Aeppli, T.\ E.\ Mason, S.\ M.\
Hayden, H.\ A.\ Mook, and J. Kulda, Science, \textbf{278}, 1432 (1997).
\bibitem{moriy} T.\ Moriya, Y.\ Takahashi, and K.\ Ueda, J.\ Phys.\ Soc.\
Jpn. \textbf{59}, 2905 (1990); A.\ J.\ Millis, H.\ Monien, and D.\ Pines,
Phys.\ Rev.\ B \textbf{42}, 167 (1990).
\bibitem{jakl} J. Jakli\v c and P. Prelov\v sek, Phys.\ Rev.\ Lett. 
\textbf{75}, 1340 (1995); Adv. Phys. \textbf{49}, 1 (2000).
\bibitem{mori} H.\ Mori, Prog.\ Theor.\ Phys.  \textbf{33}, 423 (1965).
\bibitem{sega} I.\ Sega, P.\ Prelov\v sek, and J.\ Bon\v ca, 
Phys.\ Rev.\ \textbf{68}, 054524 (2003).
\bibitem{sing} R.\ R.\ P.\ Singh and R.\ L.\ Glenister, Phys.\ Rev.\ B
\textbf{46}, 11871 (1992).
\bibitem{tyc} S.\ Ty\v c and B.\ I.\ Halperin, Phys.\ Rev.\
B\textbf{42}, 2096 (1990).
\bibitem{yosh} T.\ Yoshida {\it et al.}, Phys.\ Rev.\ Lett. \textbf{91}, 027001 (2003).

\end{thebibliography}
\end{document}